\documentclass[12pt,a4paper,preprint,superscriptaddress]{revtex4-2}

\usepackage{amsmath,amssymb}
\usepackage{hyperref}
\usepackage{orcidlink}
\usepackage{geometry} 

\begin{document}

\title{The Generalized Klein--Gordon Oscillator in Doubly Special Relativity: A Complexified Morse Interaction}

\author{Abdelmalek Boumali\,\orcidlink{0000-0003-2552-0427}}
\email{boumali.abdelmalek@gmail.com}
\affiliation{Department of Matter Science, University of Tebessa, 12000 Tebessa, Algeria}

\begin{abstract}
We investigate the one-dimensional \emph{Generalized Klein--Gordon Oscillator} (G-KGO) within Doubly Special Relativity (DSR) kinematics. The G-KGO extends the Klein--Gordon oscillator by replacing the usual linear non-minimal coupling with a general interaction function $f(x)$, leading to a factorized (SUSY-like) Schr\"odinger operator $H_-=\mathcal{A}^{\#}\mathcal{A}$ whose real spatial spectrum $\{\epsilon_n\}$ can be ensured either by Hermiticity or, for complex $f(x)$, by $\eta$-pseudo-Hermiticity and/or $\mathcal{PT}$ symmetry with a consistent metric inner product \cite{Bender1998,Bender2007RPP,BBJ2002PRL,Mostafazadeh2002,Mostafazadeh2003,Mostafazadeh2010,ElGanainy2018NatPhys}. DSR is then implemented at the level of the energy reconstruction map $\epsilon_n\mapsto E_{n,\pm}$, and we provide closed-form Magueijo--Smolin (MS) and Amelino--Camelia (AC) branches.
As an explicit and analytically controllable illustration we choose a \emph{complexified Morse interaction}: it is shape-invariant and exactly solvable \cite{Morse1929,Gendenshtein1983,Cooper1995}, it models anharmonic binding with an intrinsic finite bound-state ladder \cite{Morse1929}, and it is a standard benchmark for real spectra in non-Hermitian quantum mechanics, including imaginary-shift pseudo-Hermiticity and $\mathcal{PT}$-symmetric continuations \cite{Ahmed2001,Znojil1999,BagchiQuesne2002PLA,BagchiQuesne2002EtaOrth,Dorey2001}. We derive the Morse spatial spectrum and wavefunctions and obtain the MS/AC deformed energies after inserting the exact $\epsilon_n$. In the AC prescription an admissibility condition $\epsilon_n<4k^2$ introduces a DSR-induced truncation of the Morse ladder; we provide a compact expression for the resulting maximal bound level and discuss parameter regimes where this truncation is active. The massless limit $m=0$ sharply distinguishes the models: the MS mapping collapses to $E^2=\epsilon_n$ whereas AC remains deformed and retains the admissibility constraint.
\end{abstract}

\maketitle
\newgeometry{margin=1in} 

\section{Introduction}

Exactly solvable relativistic bound-state models remain essential for isolating how relativistic dispersion and deformation scenarios reshape spectra, eigenfunctions, and branch structure. In the scalar sector, the \emph{Klein--Gordon oscillator} (KGO) is a canonical benchmark, generated by a non-minimal coupling that produces an oscillator-like structure for the squared energies \cite{BruceMinning1993}. Its simplicity makes it particularly useful for comparative studies of relativistic deformations (including DSR) at the level of spectral reconstructions.

A natural extension replaces the linear oscillator coupling by a general function $f(x)$, yielding the \emph{Generalized Klein--Gordon Oscillator} (G-KGO). In $(1+1)$ dimensions this construction leads to a factorized Schr\"odinger-like operator $H_-=\mathcal{A}^{\#}\mathcal{A}$ whose solvability can be imported from supersymmetric quantum mechanics (SUSY QM), in particular via \emph{shape invariance} \cite{Gendenshtein1983,Cooper1995}. Allowing $f(x)$ to be complex turns the G-KGO into a controlled arena for non-Hermitian quantum mechanics: real spatial spectra can persist when the Hamiltonian is $\eta$-pseudo-Hermitian and/or $\mathcal{PT}$ symmetric, provided a suitable metric defines the physical inner product \cite{Bender1998,Bender2007RPP,BBJ2002PRL,Mostafazadeh2002,Mostafazadeh2003,Mostafazadeh2010,ElGanainy2018NatPhys}. A key mechanism is the translation-type metric $\eta=e^{-\theta p}$, which implements an imaginary coordinate shift $x\to x+i\theta$ and can map a complex potential to its complex conjugate \cite{Ahmed2001}.

On a different front, Doubly Special Relativity (DSR) introduces, in addition to $c$, a second observer-independent scale $k$ (often associated with Planckian physics) while retaining a relativity principle through deformed Lorentz transformations in momentum space \cite{AmelinoCamelia2002a,MagueijoSmolin2002,KowalskiGlikman2005,AmelinoCamelia2010,AmelinoCamelia2002Open}. Phenomenological consistency requirements (e.g.\ locality-related constraints for certain modified-dispersion scenarios) motivate treating admissibility bounds as physically meaningful rather than purely formal \cite{Hossenfelder2010PRL,Hossenfelder2013LRR}. In the present work we exploit a clean separation of layers: the spatial problem yields $\{\epsilon_n\}$, and DSR modifies only the \emph{energy reconstruction map} $\epsilon_n\mapsto E_{n,\pm}$.

\medskip
\noindent\textbf{Why a complexified Morse interaction?}
We choose a \emph{complexified Morse interaction} for three complementary reasons.

\emph{(1) Physical motivation and intrinsic finiteness.}
The Morse potential is a classic model for anharmonic molecular binding, improving on the harmonic approximation by incorporating dissociation and producing a \emph{finite} number of bound states \cite{Morse1929}. This intrinsic finiteness makes Morse an ideal testbed to disentangle two truncation mechanisms: the \emph{potential-induced} finite ladder and a \emph{DSR-induced} admissibility cutoff (present in the AC prescription).

\emph{(2) Exact solvability via shape invariance.}
Morse is shape-invariant in SUSY QM, enabling closed-form spectra and wavefunctions in terms of confluent hypergeometric/associated Laguerre functions \cite{Gendenshtein1983,Cooper1995}. This solvability persists in the G-KGO factorization, yielding analytic $\epsilon_n$ and an explicit eigenbasis.

\emph{(3) A standard non-Hermitian benchmark with real spectra.}
Complexifications of Morse are among the best studied exactly solvable non-Hermitian models with real spectra. In particular, complex Morse potentials provide textbook illustrations of pseudo-Hermiticity under imaginary coordinate shifts \cite{Ahmed2001,BagchiQuesne2002EtaOrth} and admit $\mathcal{PT}$-symmetric continuations with real, discrete spectra \cite{Znojil1999,Dorey2001,BagchiQuesne2002PLA}. This allows us to keep the spatial spectrum real while the interaction remains genuinely complex, and to isolate how DSR reshapes the final relativistic energies.

\medskip
\noindent\textbf{Main objective.}
We provide a unified, analytically controlled framework in which: (i) the G-KGO is reviewed as an exactly solvable (and, when appropriate, pseudo-Hermitian/$\mathcal{PT}$-symmetric) relativistic scalar system; (ii) MS and AC DSR deformations are implemented at the level of the energy reconstruction map; and (iii) the resulting spectral signatures are exhibited explicitly for a solvable pseudo-Hermitian complexified Morse interaction, including explicit wavefunctions, MS/AC closed forms, and the combined truncation rules.
\section{Review: the $(1+1)$ Generalized Klein--Gordon Oscillator}

\subsection{Non-minimal coupling, factorization, and Schr\"odinger-like form}

We adopt units with $c=1$ and keep $\hbar$ explicit. Consider the stationary Klein--Gordon equation
\begin{equation}
\left(p^2+m^2\right)\phi = E^2 \phi,
\qquad p=-i\hbar\frac{d}{dx}.
\label{eq:KG_free}
\end{equation}
The generalized Klein--Gordon oscillator is generated by the non-minimal substitution
\begin{equation}
p\ \longrightarrow\ \Pi \equiv p - i f(x),
\label{eq:Pi_def}
\end{equation}
where $f(x)$ is a (generally complex) interaction function. A natural scalar realization replaces $p^2$ by $\Pi^\#\Pi$:
\begin{equation}
\left(\Pi^\#\Pi + m^2\right)\phi = E^2 \phi,
\label{eq:GKGO_master}
\end{equation}
where $\#$ denotes the appropriate adjoint (standard adjoint if Hermitian; metric-adjoint if pseudo-Hermitian).

Define
\begin{equation}
\mathcal{A}=p - i f(x),\qquad
\mathcal{A}^{\#}=p + i f^{\#}(x),
\label{eq:A_Asharp}
\end{equation}
so that $\Pi=\mathcal{A}$ and $\Pi^\#=\mathcal{A}^\#$. Then \eqref{eq:GKGO_master} becomes the Schr\"odinger-like spectral problem
\begin{equation}
\mathcal{A}^{\#}\mathcal{A}\,\phi
=\left[-\hbar^2\frac{d^2}{dx^2}+V_-(x)\right]\phi
=\epsilon\,\phi,
\label{eq:schKG}
\end{equation}
with
\begin{equation}
V_-(x)=f^2(x)-\hbar f'(x),
\label{eq:Vminus}
\end{equation}
and (in the undeformed relativistic case)
\begin{equation}
\epsilon = E^2-m^2.
\label{eq:undeformed_dispersion_KG}
\end{equation}

\medskip
\noindent\textbf{Partner potential.}
The reverse ordering yields $V_+(x)=f^2(x)+\hbar f'(x)$. The two potentials are SUSY partners, but the scalar G-KGO spatial problem \eqref{eq:schKG} selects one member (here $V_-$) as the physical operator.

\subsection{SUSY viewpoint and shape invariance}

Define the partner Hamiltonians
\begin{equation}
H_-=\mathcal{A}^\#\mathcal{A},\qquad H_+=\mathcal{A}\mathcal{A}^\#,
\end{equation}
so that $H_\mp=-\hbar^2\frac{d^2}{dx^2}+V_\mp(x)$. The intertwining relations
\begin{equation}
\mathcal{A}H_- = H_+\mathcal{A},\qquad
\mathcal{A}^\# H_+ = H_- \mathcal{A}^\#
\label{eq:intertwine_KG}
\end{equation}
imply (almost) isospectrality up to the SUSY ground-state issue. If shape invariance holds,
\begin{equation}
V_+(x;a_0)=V_-(x;a_1)+R(a_0),
\qquad a_1 = f(a_0),
\label{eq:shapeinv_KG}
\end{equation}
then
\begin{equation}
\epsilon_n = \sum_{j=0}^{n-1} R(a_j),\qquad a_{j+1}=f(a_j),
\label{eq:eps_shapeinv}
\end{equation}
supplemented by normalizability \cite{Gendenshtein1983,Cooper1995}.

\subsection{$\mathcal{PT}$ symmetry and pseudo-Hermiticity}

For complex $f(x)$, $H_-$ is generally non-Hermitian in the standard inner product. Two frameworks explain why spectra may remain real.

\paragraph{(i) $\mathcal{PT}$ symmetry.}
A sufficient condition for Schr\"odinger-type operators is $V_-(x)=V_-^*(-x)$ \cite{Bender1998}.

\paragraph{(ii) Pseudo-Hermiticity.}
$H_-$ is $\eta$-pseudo-Hermitian if
\begin{equation}
H_-^\dagger=\eta\,H_-\,\eta^{-1},
\label{eq:pseudoHerm_KG}
\end{equation}
for some invertible Hermitian $\eta$. Then the physical inner product is
\begin{equation}
\langle \phi|\psi\rangle_\eta = \langle \phi|\eta|\psi\rangle,
\end{equation}
and (under standard assumptions) the spectrum is real \cite{Mostafazadeh2002,Mostafazadeh2003,Mostafazadeh2010}. A particularly useful class is the translation metric
\begin{equation}
\eta=e^{-\theta p},\qquad \eta\,x\,\eta^{-1}=x+i\theta,
\label{eq:eta_translation}
\end{equation}
which implements an imaginary shift of the coordinate and can enforce $V_-(x+i\theta)=V_-^*(x)$ \cite{Ahmed2001,BagchiQuesne2002EtaOrth}.

\subsection{Exact solution for a complexified Morse interaction in one dimension}

We now \emph{solve explicitly} the one-dimensional complexified Morse case in the G-KGO spatial problem \eqref{eq:schKG}. Consider
\begin{equation}
f(x)=D-q\,e^{-\alpha x},\qquad q\equiv A+iB,\qquad A,B,D,\alpha\in\mathbb{R}.
\label{eq:f_morse_secII}
\end{equation}
Introduce the convenient parameters
\begin{equation}
a\equiv \hbar\alpha,\qquad \lambda\equiv\frac{D}{a},\qquad q=|q|e^{i\varphi}.
\label{eq:params_morse_secII}
\end{equation}
From \eqref{eq:Vminus}, the partner potential is
\begin{align}
V_-(x)
&=f^2(x)-\hbar f'(x)
\nonumber\\
&=D^2 + q^2 e^{-2\alpha x}-q(2D+a)e^{-\alpha x}.
\label{eq:Vminus_morse_secII}
\end{align}
Hence the spatial equation \eqref{eq:schKG} becomes
\begin{equation}
\left[-\hbar^2\frac{d^2}{dx^2}+D^2 + q^2 e^{-2\alpha x}-q(2D+a)e^{-\alpha x}\right]\phi(x)=\epsilon\,\phi(x).
\label{eq:Sch_morse_secII}
\end{equation}

\paragraph{Pseudo-Hermiticity by an imaginary shift.}
Choose the shift parameter
\begin{equation}
\theta=\frac{2\varphi}{\alpha}=\frac{2}{\alpha}\tan^{-1}\!\left(\frac{B}{A}\right),\qquad \eta=e^{-\theta p}.
\label{eq:theta_eta_secII}
\end{equation}
Then $e^{-\alpha(x+i\theta)}=e^{-\alpha x}e^{-i2\varphi}$ implies
\begin{equation}
q\,e^{-\alpha(x+i\theta)}=q^*e^{-\alpha x},
\qquad\Rightarrow\qquad
f(x+i\theta)=f^*(x),
\label{eq:f_shift_secII}
\end{equation}
and consequently $V_-(x+i\theta)=V_-^*(x)$, realizing $\eta$-pseudo-Hermiticity for $H_-=-\hbar^2\partial_x^2+V_-(x)$ \cite{Ahmed2001,Mostafazadeh2010,BagchiQuesne2002EtaOrth}. Therefore, the discrete eigenvalues $\epsilon_n$ are real (under standard diagonalizability assumptions), even when $B\neq0$.

\paragraph{Reduction to the standard Morse form.}
Subtract the asymptotic constant and define $\tilde{\epsilon}\equiv \epsilon-D^2$:
\begin{equation}
\left[-\hbar^2\frac{d^2}{dx^2}+ q^2 e^{-2\alpha x}-q(2D+a)e^{-\alpha x}\right]\phi(x)=\tilde{\epsilon}\,\phi(x),
\qquad \tilde{\epsilon}<0\ \text{for bound states}.
\label{eq:Sch_morse_shifted_secII}
\end{equation}
Introduce the Morse variable
\begin{equation}
z \equiv \frac{2q}{a}\,e^{-\alpha x},
\qquad \frac{dz}{dx}=-\alpha z,
\label{eq:z_secII}
\end{equation}
and use the standard ansatz
\begin{equation}
\phi(z)=z^{s}\,e^{-z/2}\,y(z),
\qquad s>0.
\label{eq:ansatz_secII}
\end{equation}
After substitution into \eqref{eq:Sch_morse_shifted_secII}, one obtains the associated Laguerre (confluent hypergeometric) equation
\begin{equation}
z\,y''(z)+\bigl(2s+1-z\bigr)y'(z)+n\,y(z)=0,
\label{eq:LaguerreEq_secII}
\end{equation}
provided the quantization condition
\begin{equation}
s=\lambda-n,\qquad n=0,1,2,\dots
\label{eq:s_quant_secII}
\end{equation}
is satisfied. Normalizability at $x\to+\infty$ (i.e.\ $z\to0$) requires $s>0$, yielding the finite ladder
\begin{equation}
n<\lambda=\frac{D}{\hbar\alpha}.
\label{eq:nmax_secII}
\end{equation}

\paragraph{Eigenfunctions.}
With $s=\lambda-n$, the polynomial solutions of \eqref{eq:LaguerreEq_secII} are $y(z)=L_n^{(2s)}(z)$, hence
\begin{equation}
\phi_n(x)=\mathcal{N}_n\,
z^{\,\lambda-n}\,e^{-z/2}\,
L_n^{(2\lambda-2n)}(z),
\qquad
z=\frac{2q}{a}e^{-\alpha x},
\label{eq:phi_morse_secII}
\end{equation}
where $\mathcal{N}_n$ is fixed by the standard inner product in the Hermitian case ($B=0$) or by the $\eta$-inner product
$\langle\cdot|\cdot\rangle_\eta=\langle\cdot|\eta|\cdot\rangle$ in the pseudo-Hermitian case \cite{Mostafazadeh2010}.

\paragraph{Eigenvalues.}
The quantization \eqref{eq:s_quant_secII} yields the standard Morse energy shift
\begin{equation}
\tilde{\epsilon}_n=-(D-na)^2,
\label{eq:tildeeps_secII}
\end{equation}
and therefore the G-KGO separation constants are
\begin{equation}
\epsilon_n=D^2-(D-na)^2
=2D\,n a-n^2a^2
=a^2\bigl(2\lambda n-n^2\bigr),
\qquad n<\lambda.
\label{eq:eps_secII}
\end{equation}
These $\epsilon_n$ are real and independent of the complex phase of $q$ (the complexification affects eigenfunctions/metric rather than the discrete ladder), in agreement with standard complex Morse analyses \cite{Znojil1999,Ahmed2001,BagchiQuesne2002PLA}.
\section{DSR theory: kinematics and deformed dispersion}

\subsection{Conceptual overview}

DSR introduces an observer-independent energy scale $k$ while retaining a relativity principle through a deformation of Lorentz symmetry in momentum space \cite{AmelinoCamelia2002a,MagueijoSmolin2002,KowalskiGlikman2005,AmelinoCamelia2010,AmelinoCamelia2002Open}. In bound-state applications one often adopts leading-order realizations that preserve analytic tractability.

\subsection{Two prescriptions: MS and AC}

In the undeformed G-KGO, $\epsilon_n=E_n^2-m^2$. In our DSR implementations, the spatial spectrum remains $\epsilon_n$ but the energy relation is deformed.

\paragraph{MS deformation.}
\begin{equation}
E^2 - m^2\left(1-\frac{E}{k}\right)^2=\epsilon_n,
\label{eq:MS_relation_KG}
\end{equation}
which is not even in $E$ and breaks exact $E\leftrightarrow -E$ symmetry.

\paragraph{AC deformation.}
\begin{equation}
\frac{E^2-m^2}{\left(1+\frac{E}{2k}\right)^2}=\epsilon_n,
\label{eq:AC_relation_KG}
\end{equation}
which becomes singular as $\epsilon_n\to 4k^2$, motivating
\begin{equation}
\epsilon_n<4k^2.
\label{eq:AC_condition_KG}
\end{equation}

\section{G-KGO energies in DSR: closed forms and the massless limit}

\subsection{MS energies}

From \eqref{eq:MS_relation_KG} one obtains
\begin{equation}
\left(1-\frac{m^2}{k^2}\right)E^2+\frac{2m^2}{k}E-(m^2+\epsilon_n)=0,
\end{equation}
hence
\begin{equation}
E_{n,\pm}^{\text{(MS)}}=
\frac{-\frac{2m^2}{k}\pm
\sqrt{\left(\frac{2m^2}{k}\right)^2+4\left(1-\frac{m^2}{k^2}\right)(m^2+\epsilon_n)}}{2\left(1-\frac{m^2}{k^2}\right)}.
\label{eq:MS_E_KG}
\end{equation}

\subsection{AC energies and admissibility}

From \eqref{eq:AC_relation_KG} one finds
\begin{equation}
\left(1-\frac{\epsilon_n}{4k^2}\right)E^2-\frac{\epsilon_n}{k}E-(m^2+\epsilon_n)=0,
\end{equation}
hence
\begin{equation}
E_{n,\pm}^{\text{(AC)}}=
\frac{\frac{\epsilon_n}{k}\pm
\sqrt{\left(\frac{\epsilon_n}{k}\right)^2+4\left(1-\frac{\epsilon_n}{4k^2}\right)(m^2+\epsilon_n)}}{2\left(1-\frac{\epsilon_n}{4k^2}\right)}.
\label{eq:AC_E_KG}
\end{equation}
The admissibility condition is \eqref{eq:AC_condition_KG}.

\subsection{Massless limit ($m=0$)}

\paragraph{MS model.}
Setting $m=0$ in \eqref{eq:MS_relation_KG} gives
\begin{equation}
E_{n,\pm}^{\text{(MS)},\,m=0}=\pm \sqrt{\epsilon_n}.
\label{eq:MS_massless_KG}
\end{equation}

\paragraph{AC model.}
Setting $m=0$ in \eqref{eq:AC_relation_KG} yields, with $s_n=\sqrt{\epsilon_n}$,
\begin{equation}
E_{n,+}^{\text{(AC)},\,m=0}=\frac{2k\,s_n}{2k-s_n},
\qquad
E_{n,-}^{\text{(AC)},\,m=0}=-\frac{2k\,s_n}{2k+s_n},
\qquad s_n<2k.
\label{eq:AC_massless_KG}
\end{equation}

\section{Example: a pseudo-Hermitian complexified Morse interaction}

\subsection{Interaction choice, complexification, and pseudo-Hermiticity mechanism}

We adopt the complexified Morse-type interaction
\begin{equation}
f(x)=D-(A+iB)e^{-\alpha x},
\qquad A,B,D,\alpha\in\mathbb{R},
\label{eq:morse_f_KG}
\end{equation}
and define
\begin{equation}
q\equiv A+iB = |q|e^{i\varphi},\qquad a\equiv \hbar\alpha,\qquad \lambda\equiv \frac{D}{a}.
\label{eq:shorthand_V}
\end{equation}
The dependence on the \emph{phase} of $q$ can be absorbed by an imaginary shift of $x$, which is the pseudo-Hermiticity mechanism emphasized in \cite{Ahmed2001,Mostafazadeh2003,Mostafazadeh2010,BagchiQuesne2002EtaOrth}. Choose
\begin{equation}
\theta=\frac{2\varphi}{\alpha}=\frac{2}{\alpha}\tan^{-1}\!\left(\frac{B}{A}\right),
\qquad
\eta=e^{-\theta p},
\label{eq:theta_eta}
\end{equation}
so that $e^{-\alpha(x+i\theta)}=e^{-\alpha x}e^{-i2\varphi}$ and therefore
\begin{equation}
q\,e^{-\alpha(x+i\theta)}=q^*\,e^{-\alpha x}.
\end{equation}
Hence
\begin{equation}
f(x+i\theta)=D-q\,e^{-\alpha(x+i\theta)} = D-q^*e^{-\alpha x}=f^*(x),
\label{eq:f_shift_conj}
\end{equation}
and the induced potential satisfies $V_-(x+i\theta)=V_-^*(x)$, realizing $\eta$-pseudo-Hermiticity for $H_-=-\hbar^2\partial_x^2+V_-(x)$ \cite{Ahmed2001,Mostafazadeh2003,Mostafazadeh2010,BagchiQuesne2002EtaOrth}.

\subsection{Partner potential and reduction to a standard Morse problem}

With \eqref{eq:morse_f_KG} the potential \eqref{eq:Vminus} becomes
\begin{align}
V_-(x)
&= f^2(x)-\hbar f'(x)
\nonumber\\
&=D^2 + q^2 e^{-2\alpha x}-q(2D+a)e^{-\alpha x}.
\label{eq:Vminus_morse}
\end{align}
The eigenvalue equation \eqref{eq:schKG} is
\begin{equation}
\left[-\hbar^2\frac{d^2}{dx^2}+D^2 + q^2 e^{-2\alpha x}-q(2D+a)e^{-\alpha x}\right]\phi=\epsilon\,\phi.
\label{eq:Sch_V}
\end{equation}
It is convenient to subtract the asymptotic constant $D^2$ and define
\begin{equation}
\tilde{\epsilon}\equiv \epsilon-D^2,
\qquad
\left[-\hbar^2\frac{d^2}{dx^2}+ q^2 e^{-2\alpha x}-q(2D+a)e^{-\alpha x}\right]\phi=\tilde{\epsilon}\,\phi.
\label{eq:Sch_shifted}
\end{equation}
Bound states correspond to $\tilde{\epsilon}<0$ and the quantization is finite. The complex Morse structure is a standard solvable non-Hermitian benchmark \cite{Znojil1999,Dorey2001,BagchiQuesne2002PLA}.

\subsection{Spatial spectrum: exact quantization and finite bound-state count}

Introduce the Morse variable
\begin{equation}
z \equiv \frac{2q}{a}\,e^{-\alpha x},
\qquad \frac{dz}{dx}=-\alpha z.
\label{eq:z_def_V}
\end{equation}
The standard Morse reduction yields the quantization condition
\begin{equation}
\tilde{\epsilon}_n=-(D-na)^2,
\qquad n=0,1,2,\dots
\label{eq:tildeeps}
\end{equation}
and therefore
\begin{equation}
\epsilon_n = D^2-(D-na)^2
=2D\,n a-n^2 a^2
=a^2\bigl(2\lambda n-n^2\bigr).
\label{eq:eps_morse_KG}
\end{equation}
Normalizability yields a \emph{finite} ladder:
\begin{equation}
n<\lambda=\frac{D}{\hbar\alpha}.
\label{eq:nmax_morse_KG}
\end{equation}
This intrinsic finiteness is the Morse hallmark \cite{Morse1929} and will compete with the AC admissibility cutoff.

\medskip
\noindent\textbf{Monotonicity.}
For $0\le n<\lambda$,
\begin{equation}
\epsilon_{n+1}-\epsilon_n=a^2\bigl(2\lambda-(2n+1)\bigr)>0,
\end{equation}
so $\epsilon_n$ increases monotonically with $n$ across the bound ladder.

\subsection{Morse wavefunctions: Laguerre form and inner-product issues}

A convenient Laguerre representation of the bound states is
\begin{equation}
\phi_n(x)=\mathcal{N}_n\,
z^{\,\lambda-n}\,e^{-z/2}\,
L_n^{(2\lambda-2n)}(z),
\qquad
z=\frac{2q}{a}e^{-\alpha x},
\label{eq:morse_wf_V}
\end{equation}
where $L_n^{(\nu)}$ are associated Laguerre polynomials. In the Hermitian Morse case ($B=0$ and $q>0$), normalization may be taken in the standard inner product. In the pseudo-Hermitian complex case ($B\neq 0$), physical orthogonality and normalization are defined with the $\eta$-inner product
\begin{equation}
\langle \phi_m|\phi_n\rangle_\eta
=\int_{-\infty}^{\infty}\phi_m^*(x)\,(\eta\phi_n)(x)\,dx,
\qquad \eta=e^{-\theta p},
\label{eq:eta_inner_V}
\end{equation}
which is the natural setting when $H_-$ is $\eta$-pseudo-Hermitian \cite{Mostafazadeh2010,BagchiQuesne2002EtaOrth}.

\subsection{Undeformed Klein--Gordon energies for the Morse spectrum}

In the undeformed relativistic case, \eqref{eq:undeformed_dispersion_KG} gives
\begin{equation}
E_{n,\pm}^{(0)}
=\pm\sqrt{m^2+\epsilon_n}
=\pm\sqrt{m^2+a^2(2\lambda n-n^2)}.
\label{eq:KG_undeformed_morse}
\end{equation}

\subsection{MS energies for the Morse spectrum: explicit branches and constraints}

Substituting \eqref{eq:eps_morse_KG} into \eqref{eq:MS_E_KG} yields
\begin{equation}
E_{n,\pm}^{\text{(MS)}}=
\frac{-\frac{2m^2}{k}\pm
\sqrt{\left(\frac{2m^2}{k}\right)^2
+4\left(1-\frac{m^2}{k^2}\right)\left[m^2+a^2(2\lambda n-n^2)\right]}}
{2\left(1-\frac{m^2}{k^2}\right)}.
\label{eq:MS_morse_explicit}
\end{equation}
Because MS is not even in $E$, the positive and negative branches are not symmetric at finite $k$. The Morse bound-state count remains \eqref{eq:nmax_morse_KG}; in addition one typically assumes $k>m$ so that $1-m^2/k^2>0$.

\subsection{AC energies for the Morse spectrum: criticality and DSR truncation}

Substituting \eqref{eq:eps_morse_KG} into \eqref{eq:AC_E_KG} yields
\begin{equation}
E_{n,\pm}^{\text{(AC)}}=
\frac{\frac{a^2(2\lambda n-n^2)}{k}\pm
\sqrt{\left(\frac{a^2(2\lambda n-n^2)}{k}\right)^2
+4\left(1-\frac{a^2(2\lambda n-n^2)}{4k^2}\right)\left[m^2+a^2(2\lambda n-n^2)\right]}}
{2\left(1-\frac{a^2(2\lambda n-n^2)}{4k^2}\right)}.
\label{eq:AC_morse_explicit}
\end{equation}
The AC admissibility condition \eqref{eq:AC_condition_KG} becomes
\begin{equation}
a^2(2\lambda n-n^2)<4k^2.
\label{eq:AC_morse_adm}
\end{equation}
Since $\epsilon_n$ increases with $n$ for $n<\lambda$, the AC constraint truncates the \emph{highest} Morse levels when it is active. Solving $\epsilon_n=4k^2$ gives the root in $[0,\lambda]$
\begin{equation}
n_\star=\lambda-\sqrt{\lambda^2-\left(\frac{2k}{a}\right)^2}
=\frac{D-\sqrt{D^2-4k^2}}{\hbar\alpha},
\qquad (D>2k),
\label{eq:nstar_AC}
\end{equation}
so the admissible bound states satisfy
\begin{equation}
n=0,1,2,\dots,\qquad
n<\min\!\left(\lambda,\;n_\star\right).
\label{eq:nmax_total_AC}
\end{equation}
If $D\le 2k$ then $\epsilon_n\le D^2\le 4k^2$ for all Morse bound states, and AC does not further truncate the ladder.

\subsection{Massless limit for Morse: MS collapse vs AC persistence}

For $m=0$, MS collapses to the undeformed map:
\begin{equation}
E_{n,\pm}^{\text{(MS)},\,m=0}=\pm\sqrt{\epsilon_n}
=\pm a\,\sqrt{2\lambda n-n^2}.
\label{eq:MS_morse_massless}
\end{equation}
In contrast, AC remains deformed. With $s_n\equiv \sqrt{\epsilon_n}=a\sqrt{2\lambda n-n^2}$,
\begin{equation}
E_{n,+}^{\text{(AC)},\,m=0}=\frac{2k\,s_n}{2k-s_n},
\qquad
E_{n,-}^{\text{(AC)},\,m=0}=-\frac{2k\,s_n}{2k+s_n},
\qquad s_n<2k,
\label{eq:AC_morse_massless}
\end{equation}
and $s_n<2k$ is exactly \eqref{eq:AC_morse_adm}.

\section{Discussion}

The complexified Morse interaction provides a transparent analytic benchmark for the G-KGO in DSR because (i) the full spatial problem is solvable in closed form by shape invariance \cite{Gendenshtein1983,Cooper1995}, (ii) the discrete spatial spectrum $\epsilon_n$ is real and intrinsically finite \cite{Morse1929}, and (iii) non-Hermiticity is controlled by a constructive pseudo-Hermitian mechanism based on an imaginary shift of the coordinate \cite{Ahmed2001,Mostafazadeh2003,Mostafazadeh2010,BagchiQuesne2002EtaOrth}. In this setting the complexification parameters $(A,B)$ primarily reshape eigenfunctions and the metric, while leaving the discrete ladder \eqref{eq:eps_morse_KG} invariant; this separates non-Hermitian structure (inner product and eigenbasis) from kinematical deformation (energy reconstruction).

DSR then acts purely algebraically. In the MS prescription the deformation yields branch-asymmetric spectra through the non-even dependence on $E$ in \eqref{eq:MS_relation_KG}, while the Morse finiteness \eqref{eq:nmax_morse_KG} remains the universal truncation mechanism. In contrast, the AC prescription introduces a criticality at $\epsilon_n\to 4k^2$, producing the additional admissibility inequality \eqref{eq:AC_morse_adm}. Because the Morse ladder is monotone in $n$, this AC condition truncates the \emph{highest} bound levels whenever $D>2k$, leading to the combined cutoff summarized by \eqref{eq:nmax_total_AC}. The massless limit highlights the qualitative difference: MS collapses to the undeformed mapping $E^2=\epsilon_n$, whereas AC retains a nonlinear map and the same admissibility threshold.

From a phenomenological perspective, DSR-inspired modified dispersion relations are also constrained by locality and related consistency considerations, which motivates treating admissibility bounds (such as $\epsilon_n<4k^2$ in AC) as physically meaningful rather than purely formal \cite{Hossenfelder2010PRL,Hossenfelder2013LRR,AmelinoCamelia2002Open}. More broadly, complexified solvable families such as Morse remain valuable because they permit controlled non-Hermitian effects without sacrificing analytic tractability \cite{Bender2007RPP,ElGanainy2018NatPhys}.

\section{Conclusion and outlook}

We reviewed the $(1+1)$ generalized Klein--Gordon oscillator as a factorized (SUSY-like) relativistic scalar system and discussed $\mathcal{PT}$ symmetry/pseudo-Hermiticity as mechanisms ensuring spectral reality for complex interactions. For a complexified Morse interaction we derived the spatial spectrum and wavefunctions, exhibited an explicit pseudo-Hermitian imaginary-shift metric, and obtained MS/AC DSR-deformed energies in closed form. AC enforces an additional admissibility cutoff $\epsilon_n<4k^2$ which can further truncate the intrinsically finite Morse ladder when $D>2k$, while MS does not introduce an analogous universal truncation in the present realization. Extensions include higher-dimensional G-KGOs, external fields, and thermodynamic analyses based on the deformed spectra.


\end{document}